\documentclass[prb,twocolumn,showpacs,amsmath,amssymb]{revtex4}
\usepackage{graphicx} 
\usepackage{dcolumn}  
\usepackage{bm}       
\begin{document}
\title{Saturation of Two Level Systems and
Charge Noise in Josephson Junction Qubits}
\author{Magdalena Constantin,$^{1,\dagger}$ 
Clare C. Yu,$^1$ and John M. Martinis$^2$}
\affiliation{
$^1$ Department of Physics and Astronomy, University of California,
Irvine, California 92697\\
$^2$ Department of Physics, University of California, Santa Barbara,
California 93106}
\date{\today}
\begin{abstract}
We study the charge noise $S_Q$ in Josephson qubits produced
by fluctuating two level systems (TLS) with electric dipole moments 
in the substrate. The TLS are driven 
by an alternating electric field of angular frequency $\Omega$ and 
electric field intensity $I$. It is not widely appreciated that TLS 
in small qubits can easily be strongly saturated if $I\gg I_c$, where 
$I_c$ is the critical electric field intensity. To investigate the effect
of saturation on the charge noise, we express the noise spectral 
density in terms of density matrix elements. To determine the
dependence of the density matrix elements on the ratio $I/I_c$, 
we find the steady state 
solution for the density matrix using the Bloch-Redfield differential 
equations.  We then obtain an expression for the spectral 
density of charge fluctuations as a function of frequency $f$ and 
the ratio $I/I_c$. We find $1/f$ charge noise at low frequencies, 
and that the charge noise is white (constant) at high frequencies. 
Using a flat density of states,
we find that TLS saturation has no effect on the charge noise at either high
or low frequencies.
\end{abstract}

\pacs{74.40.+k, 03.65.Yz, 03.67.-a, 85.25.-j}
\maketitle

\section{Introduction}
The superconducting Josephson junction qubit is a leading candidate
in the design of a quantum computer, with several experiments
demonstrating single qubit preparation, manipulation,
and measurement, \cite{Vion2002,Yu2002,Martinis2002,Chiorescu2003}
as well as the coupling of qubits. \cite{Pashkin2003,Berkley2003}
A significant advantage of this
approach is scalability, as these qubits may be readily fabricated in
large numbers using integrated circuit technology. However,
noise and decoherence are major obstacles to using superconducting
Josephson junction qubits to construct quantum computers. Recent
experiments \cite{Simmonds2004,Martinis2005} indicate that a dominant
source of decoherence is two level systems (TLS) that are fluctuating
in the insulating barrier of the tunnel junction as well as in the 
dielectric material used to fabricate the circuit, e.g., the substrate. 
The two level fluctuators that have electric dipole moments can induce
image charges in the nearby superconductor and hence 
produce charge noise $S_Q(\omega)$. 
\cite{Martinis1992,Mooij1995,Zorin1996,Kenyon2000,Astafiev2006,Shnirman2005,Faoro2005,Faoro2006}

Previous theories of charge noise\cite{Shnirman2005,Faoro2005,Faoro2006} 
have neglected the important issue of the saturation of the two level systems by 
electric fields used to manipulate the qubits. Dielectric (ultrasonic)
experiments on insulating glasses at low temperatures have found that
when the electric (acoustic) field intensity $I$ used to make the 
measurements exceeds the critical intensity $I_c$, the dielectric (ultrasonic) 
power absorption by the TLS is saturated, and the attenuation decreases as
the field intensity increases.
\cite{Arnold1976,Golding1976,Graebner1983,Schickfus1977,Martinis2005}
(If ${\cal{E}}\cos(\Omega t)$ denotes the electric field, then we define
the intensity $I={\cal{E}}^{2}$.) 
Previous theories of charge noise in Josephson junctions assumed
that the TLS were not saturated, i.e., that $I\ll I_c$. This seems 
sensible since charge noise experiments \cite{Astafiev2004} have been 
done in the limit where the qubit absorbed only one photon. 

\begin{figure}
\includegraphics[height=4.2cm,width=6.5cm]{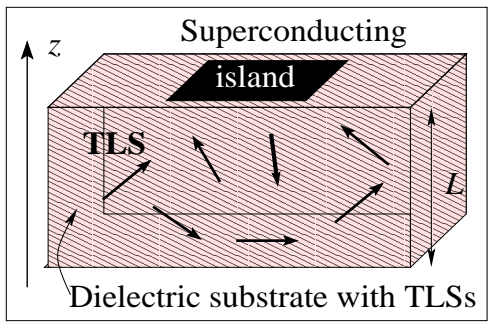}
\caption{\label{fig1} (Color online)
Distribution of two level systems (TLS) in the 
dielectric substrate of a Cooper pair box. The superconducting island 
with the Josephson junction qubit is the black rectangle on the top 
of the substrate. There is a ground plane underneath the substrate.}
\end{figure}

However, the following simple estimate shows that
stray electric fields associated with this photon
could saturate two level systems in the dielectric substrate
which supports the qubit. We can estimate the voltage $V$
across the capacitor associated with the substrate and ground plane
beneath a Cooper pair box (see Fig.~\ref{fig1}) by setting 
$CV^2/2=\hbar\omega$ where $\hbar\omega$ is the energy of the
microwave photon. We estimate the capacitance
$C=\varepsilon_o\varepsilon_r A/L \sim 7$ aF using the area
$A=40\times 800$ nm$^2$ of the Cooper pair box,
the thickness $L=400$ nm of the substrate,\cite{Astafiev2004} and
the relative permittivity $\varepsilon_r=10$. Using $\omega/2\pi=f=10$
GHz, we obtain a voltage of $V\sim 1.4\;$mV. A substrate
thickness $L$ of 400 nm yields an electric field of
${\cal{E}}\sim 3.4\times 10^{3}$ V/m. 

We can compare this with
the critical intensity $I_c$ and the associated
critical electric field ${\cal{E}}_c$ which has been measured
experimentally \cite{Martinis2005}. 
Martinis {\it et al.} \cite{Martinis2005} measured the low temperature 
dielectric loss tangent of
amorphous SiO$_2$ at $f=7.2$ GHz and amorphous SiN$_{x}$ at $f=4.7$ GHz 
as a function of the root-mean-square (rms) voltage. They found
that the loss tangent was constant at low power, but rolled over and
decreased above a critical rms voltage $V_{c} \sim 0.2\;\mu$V.
For a capacitor thickness of 300 nm, the associated critical field is
${\cal{E}}_c\sim 0.7$ V/m. So
${\cal{E}}/{\cal{E}}_{c}\sim 5\times 10^{3}$, and
$I/I_{c}=\left({\cal{E}}/{\cal{E}}_{c}\right)^{2}\sim 2\times 10^{7}\gg 1$.

We can do a similar estimate to show that a single photon would
even more strongly saturate resonant TLS in the insulating barrier
of the tunnel junction. We use the same numbers as before but with
$C=1$ fF and the thickness of the junction $L=1.5$ nm to
obtain ${\cal{E}}\sim 7\times 10^4$ V, and $I/I_{c}\sim 10^{10}\gg 1$.

However, there are only a few TLS in the oxide barrier of a
small tunnel junction as the following estimate
shows. For a parallel plate capacitor with
$L=1.5$ nm and $A=1\;\mu$m$^2$, 
the volume is $V_o=1.5\times 10^{-21}$ m$^3$.
The typical TLS density of states \cite{Phillips} is $10^{46}/Jm^{3}$.
However, only a fraction of these have electric dipole moments. So we
will assume that the density of states of TLS with electric dipole moments 
is $P_{TLS}\simeq 10^{45}/Jm^{3}\simeq 663/h$GHz$\mu$m$^3$. Using this
value of $P_{TLS}$, we find that in a small tunnel junction,
there are only 2 TLS with an energy splitting less than 10 GHz. 
A single fluctuator would have a Lorentzian noise spectrum. 
The presence of $1/f$ noise implies many more than 2 fluctuators. 
It is likely that these additional fluctuators are in the substrate.
Our main point is that TLS in small devices are easily saturated.
It is therefore important to analyze the effect of TLS saturation on the
charge noise both at low and high frequencies $f$ of the noise spectrum.

In this article, we explore the consequences of this saturation on the 
spectral density of polarization and charge fluctuations. We consider a 
driven system consisting of two level systems with electric dipole 
moments that fluctuate randomly, leading to fluctuations $\delta P(t)$ 
in the polarization denoted by $P(t)$. 
In addition, the dipole moments of these TLS couple 
to an applied ac electric field that drives the system with an angular 
frequency $\Omega$. 

Let us consider a single two level system. 
According to the Wiener-Khintchine theorem, in the stationary state 
the polarization noise spectral density $S_P(\omega)$ of this
two level system is twice the Fourier
transform of the polarization autocorrelation function: 
\begin{eqnarray}
S_P(\omega) &=& 2\int_{-\infty}^{+\infty}\mbox{d}(t_1-t_2)
\mbox{e}^{i\omega (t_1-t_2)} 
\left[\langle \delta P(t_1)\delta P(t_2) \rangle\right]_{\rho}\nonumber\\
&=& 2\int_{-\infty}^{+\infty}\mbox{d}(t_1-t_2)\mbox{e}^{i\omega (t_1-t_2)}
\mbox{Tr}\{ \rho \langle\delta P(t_1) \delta P(t_2)\rangle\}\nonumber
\end{eqnarray}
where $[...]_{\rho}$ denotes a trace over the density matrix $\rho$,
and $\langle ... \rangle$ denotes an average over the time series. 
$\delta P(t) = [P(t) - \langle P\rangle]$ is the fluctuation in the
polarization $P(t)$ at time $t$ and $\langle P\rangle$ is the
polarization averaged over the time series. 
From the convolution theorem, $S_P(\omega)\sim |\delta P(\omega)|^2$.
If the average over the density matrix is equivalent to an average over
the time series, then one can easily show that $\delta P(\omega=0)=0$
and hence, $S_P(\omega=0)=0$.

We assume that the total density matrix $\rho_T$ is the product of two factors:
one factor $\rho$ that contains the contributions of the ac driving field and the 
other that contains the contributions of the random fluctuations of the dipoles.
We make this division because we can find the time dependence of the
density matrix (in the Schroedinger representation)
due to the driving field by solving the Bloch-Redfield
equations. Since it is not clear how to include the random fluctuations of
the two level systems in the Hamiltonian and hence, in the Bloch-Redfield
equations, we will treat the random fluctuations separately. So
the density matrix $\rho$ contains the contributions
of the driving field to $S_P(\omega)$ while the polarization autocorrelation
function $\langle \delta P(t_1) \delta P(t_2) \rangle$ contains the
contributions of the random fluctuations of the electric dipoles.
We will solve the Bloch-Redfield equations in steady state to find 
the time evolution of the density matrix $\rho(t)$ and its dependence 
on the ratio $I/I_c(\Omega,T)$ of the electric field intensity $I$ to the 
critical intensity $I_c(\Omega,T)$ which is a function of the driving angular 
frequency $\Omega$ and the temperature $T$. The benefit of this approach is that
it is valid in both equilibrium as well as steady state non-equilibrium 
situations. From the polarization noise spectral density, we can obtain the charge
noise $S_Q(\omega)$. We then average over the distribution of independent TLS.
Unlike previous theoretical efforts,\cite{Shnirman2005,Faoro2005,Faoro2006}
we use the standard TLS density of states that is a constant independent of energy.

At low frequencies ($hf \ll k_B T$) the system is in equilibrium,
and we find $1/f$ charge noise that is
proportional to the temperature and to the dielectric loss tangent 
$\tan\delta$ that has a well known contribution from the electric dipole moments
of TLS.\cite{Phillips,Classen1994,Arnold1976} 
In addition the low frequency charge noise has negligible dependence on
the electric field intensity ratio $I/I_c$.

At high frequencies ($h f \gg k_B T$) we find that the charge noise is 
white noise independent of frequency. It has a very weak dependence on the 
ratio $I/I_c(\Omega,T)$ and the driving frequency $\Omega$.  
We also find that the amplitude of the high frequency white charge 
noise decreases gradually as the temperature increases. The fact 
that the charge noise spectrum depends very weakly on the ratio 
$I/I_c(\Omega,T)$ indicates that the saturation of two level systems 
does not affect charge noise. 

The paper is organized as follows. In Section \ref{Model}, we present our
model of a TLS in an external driving field. 
In Section \ref{FDT}, we use the fluctuation-dissipation theorem to give an
expression for the charge noise in thermal equilibrium. 
In Section \ref{Noise}, we take a more general approach that is valid
in both equilibrium and nonequilibrium cases. In particular,
we derive a general analytic expression for the spectral 
density of polarization and charge fluctuations of an individual two level 
system (also referred to as a fluctuator) in terms of the density matrix. 
In Section \ref{B-R}, we solve the Bloch-Redfield linear differential
equations for the density matrix. We find the steady state
solution of the Bloch-Redfield equations and
we analyze its dependence on the ratio $I/I_c(\Omega,T)$.
In Section \ref{Results}, we investigate the noise spectrum of a 
single random telegraph fluctuator. We then average over the distribution 
of independent TLS numerically to determine the frequency dependence of 
the noise spectrum. A summary is given in Section \ref{Concl}.

\section{Two Level System (TLS)}
\label{Model}

In applying the standard model of two level systems to Josephson
junction devices, we consider a TLS that sits in the insulating
substrate or in the tunnel barrier, and has an electric dipole moment
${\bf p}$ consisting of a pair of opposite charges separated by a
distance $d$. The electrodes positioned at $z=0$ and $z=L$ are kept 
at the same potential. The angle between ${\bf p}$ and the $z$--axis, 
perpendicular to the electrodes, is $\theta$. The dipole induces
charge on the electrodes. As we show in the appendix, the magnitude of the 
induced charge $Q$ on each electrode is proportional to the
$z$-component of the dipole moment, $p_z=p\cos \theta$, i.e.,
\begin{equation}
Q=\left|\frac{p\cos\theta}{L}\right| 
\label{eq:Q}
\end{equation}
The dipole flips
and induces polarization fluctuations and hence charge fluctuations
on the electrodes.

The TLS is in a double--well potential with a tunneling matrix
element $\Delta_0$ and an asymmetry energy $\Delta$.\cite{Phillips}
The Hamiltonian of a TLS in an external ac field can be written as
$H(t)=H_0+H_1(t)$, where $H_0=\frac{1}{2}(\Delta \sigma_z+\Delta_0 \sigma_x)$,
and $H_1(t)=-\sigma_z {\bf p}\cdot\bm{\xi}_{ac}(t)$. Here
$\sigma_{x,z}$ are the Pauli spin matrices and
$\bm {\xi}_{ac}(t)=\bm {\xi}_{ac}\cos\Omega t$ is a small
perturbing ac electric field of angular frequency $\Omega$ that couples to the
TLS electric dipole moment. $\bm {\xi}_{ac}$ points along the $z-$axis.

After diagonalization of $H_0$, the Hamiltonian becomes
\begin{align}
H(t) &= H_0 + H_1(t)\label{diagH} \\
H_0&= \frac{1}{2} E\sigma_z \label{diagH0} \\
H_1(t)& = -\eta(\Delta_0 \sigma_x + \Delta \sigma_z)\mbox{cos}\Omega t,
\label{eq:diagH1}
\end{align}
where $E=\sqrt{\Delta^2 + \Delta_{o}^2}$ is the TLS energy splitting
and $\eta \equiv {\bf p}\cdot\bm{\xi}_{ac}/E$. Notice that $\eta$ is a
small dimensionless variable ($\eta\approx 5\times 10^{-3}$ for
$p=3.7$ D, $\xi_{ac}\approx 10^3$ V/m and $E/h \approx 10$ GHz).
(The dipole moment of an OH$^{-}$ molecule is 3.7 D.\cite{Golding1979})
The energy eigenbasis is denoted by $\{|+\rangle, |-\rangle \}$, and 
the corresponding eigenvalues are $E_{\pm}=\pm E/2$, where 
$+$ $(-)$ refers to the upper (lower) level of the TLS.
The energy splitting $E$ will also be referred to as $\hbar \omega_0$.

An excited two level system can decay to the ground state by emitting
a phonon. The longitudinal relaxation rate is given by:\cite{Phillips}
\begin{equation}
T_1^{-1}=\frac{\gamma_d^2}{\left(2\pi\rho\hbar^4\right)}
\left[\left(\frac{1}{c_{\ell}^{5}}\right)+\left(\frac{2}{c_{t}^{5}}\right)\right]
E \Delta_o^{2}\coth\left(\frac{E}{2k_BT}\right)~,
\label{eq:T1inv}
\end{equation}
where $\rho$ is the mass density, $c_{\ell}$ is the longitudinal
speed of sound, $c_{t}$ is the transverse speed of sound, and
$\gamma_d$ is the deformation potential. In this paper we will use
the values for SiO$_{2}$:
$\gamma_d=$1 eV, $\rho=$ 2200 kg/m$^{3}$, $c_{\ell}$=5800
m/s, and $c_{t}$=3800 m/s. Typically, $T_1$ varies between
$10^{-9}$ s and $10^{4}$ s for temperatures around 0.1 K.
The distribution of TLS parameters is given by \cite{Phillips1987,Phillips}
\begin{equation}
P(E,T_1)=\frac{P_{TLS}}{2T_1\sqrt{1-\left(\tau_{min}(E)/T_1\right)}},
\label{eq:PET1}
\end{equation}
where $P_{TLS}$ is a constant density of states that represents the
number of TLS per unit energy and unit volume. The minimum
relaxation time $\tau_{min}(E)$ corresponds to $T_1$ for a symmetric
double--well potential (i.e., $E=\Delta_0$):
\begin{equation}
\tau_{min}^{-1}=\frac{\gamma_d^2}{\left(2\pi\rho\hbar^4\right)}
\left[\left(\frac{1}{c_{\ell}^{5}}\right)+\left(\frac{2}{c_{t}^{5}}\right)\right]
E^{3}\coth\left(\frac{E}{2k_BT}\right) 
\end{equation}
Alternatively, the TLS distribution function can be expressed in terms 
of the TLS matrix elements $\Delta$
and $\Delta_0$: 
\begin{equation}
P(\Delta,\Delta_0)=\frac{P_{TLS}}{\Delta_0}.
\label{eq:PTLS}
\end{equation} 
The typical range of
values for $\Delta$ and $\Delta_0$ are $0 \leq \Delta/k_B \leq 4$ K and
2 $\mu$K $\leq \Delta_0/k_B \leq 4$ K, where $k_B$ is the
Boltzmann's constant. We will use these values for our numerical 
integrations in Section \ref{Results}.

\section{Thermal Equilibrium Expression for Charge Noise}
\label{FDT}
We begin by considering the case of thermal equilibrium. 
According to the Wiener-Khintchine theorem, the charge spectral density 
$S_Q(\omega)$ is twice the Fourier transform $\Psi_Q(\omega)$ 
of the autocorrelation function of the fluctuations in the charge.
In equilibrium we can use the fluctuation-dissipation theorem\cite{Forster}
to find that the (unsymmetrized) charge noise is given by:
\begin{equation}
S_{Q}(\bm{k},\omega)=\frac{4\hbar}{1-\mbox{e}^{-\hbar \omega/k_BT}}
\chi_{Q}^{\prime\prime}(\bm{k},\omega),
\label{eq:S_Q}
\end{equation}
where $Q$ is the induced (bound) charge on the electrodes and 
$\chi_{Q}^{\prime\prime}(\bm{k},\omega)$ is the Fourier transform of
\begin{equation}
\chi_{Q}^{\prime\prime}(\bm{r},t;\bm{r}^{\prime},t^{\prime})=
\frac{\langle\left[Q(\bm{r},t),Q(\bm{r}^{\prime},t^{\prime})\right]_{c}\rangle_{e}}{2\hbar}
\end{equation}
where $[...]_c$ is a commutator, and $\langle ... \rangle_{e}$ is an ensemble average.
We use $Q=\int \bm{P}\cdot d\bm{A}$, where $\bm{P}$ is the electric 
polarization density, and choose 
$P_z$ and $d\bm{A} \| \hat{z}$ since $Q\sim |p_z|$ to find
\begin{equation}
\chi_{Q}^{\prime\prime}(\bm{k},\omega)=\varepsilon_{o}A^2
\chi_{P_z}^{\prime\prime}(\bm{k},\omega)
\end{equation} 
where $\varepsilon_0$ is the vacuum permittivity, $A$ is the area of a 
plate of the parallel plate capacitor with capacitance $C$, and 
$\chi_{P_z}^{\prime\prime}(\bm{k},\omega)$ is the imaginary part 
of the electric susceptibility. We set $\bm{k}=0$, and use
\begin{equation}
\varepsilon_o\chi_{P_z}^{\prime\prime}(\omega)=
\varepsilon^{\prime}(\omega)\tan\delta(\omega)
\end{equation} 
where the dielectric loss tangent
$\tan\delta(\omega)=\varepsilon^{\prime\prime}(\omega)/\varepsilon^{\prime}(\omega)$.
$\varepsilon^{\prime}(\omega)$ and $\varepsilon^{\prime\prime}(\omega)$
are the real and imaginary parts of the dielectric permittivity, respectively.
We also use 
\begin{equation}
C=\varepsilon^{\prime}A/L
\label{eq:Cap}
\end{equation} 
to find
\begin{equation}
S_Q(\omega)=\frac{4\hbar C}{1-\mbox{e}^{-\hbar \omega/k_BT}} 
\tan\delta(\omega),
\label{eq:S_Q_Tan_delta}
\end{equation}
where $S_Q(\omega)\equiv S_Q(\bm{k}=0,\omega)/V_o$, the volume of 
the capacitor is $V_o=AL$, and $\varepsilon^{\prime}(\omega)=
\varepsilon^{\prime}+\varepsilon_{TLS}(\omega)\simeq\varepsilon^{\prime}=
\varepsilon_0 \varepsilon_r$ where $\varepsilon_r$ is the relative permittivity.
The frequency dependent permittivity $\varepsilon_{TLS}(\omega)$ 
produced by TLS is negligible 
compared to the constant permittivity $\varepsilon^{\prime}$.\cite{Phillips} 
The TLS dynamic electric susceptibilities ($\chi^{\prime}(\omega)$,
$\chi^{\prime\prime}(\omega)$), and hence the dielectric loss 
tangent, can be obtained by solving the Bloch equations in 
equilibrium.\cite{Jackle1975,Arnold1976,Graebner1983} One
can then average over the distribution of TLS parameters. However, since we 
will be considering driven systems that are in a nonequilibrium steady
state, we need to take a more general approach which is described in
the next section.

\section{General Expression for Spectral Density of Polarization 
and Charge Fluctuations of a Two Level System}
\label{Noise}
The noise in our model is due to a
fluctuating two level system with an electric dipole moment
that changes its orientation with respect to the direction of the
applied driving field while keeping its magnitude constant.
In this section, we begin by deriving a general expression for the polarization 
noise $S_P(\omega)$ of a single TLS that is valid at all frequencies
and in both equilibrium and nonequilibrium situations. Since we are interested
in TLS saturation, this formulation will apply to a driven system
in nonequilibrium steady state. We then relate the polarization noise to
the charge noise $S_Q(\omega)$. 

According to the Wiener-Khintchine theorem, the polarization spectral density
$S_P(\omega)$ in the stationary state is twice the Fourier transform of the 
autocorrelation function of the fluctuations in the polarization:
\begin{equation}
S_P(\omega) = 
2\int_{-\infty}^{+\infty}\mbox{d}(t_1-t_2)\mbox{e}^{i\omega (t_1-t_2)}
\left[\langle P_H(t_1) P_H(t_2) \rangle - \langle P \rangle^2 \right]_{\rho},
\label{eq:Pnoise}
\end{equation}
where the subscript $H$ denotes the Heisenberg representation and 
$\langle P\rangle$ is the time averaged value which is 
independent of the actual representation. We can rewrite 
this expression in the Heisenberg representation as
\begin{align}
S_P(\omega) &= 
2\int_{-\infty}^{+\infty}\mbox{d}(t_1-t_2)\mbox{e}^{i\omega (t_1-t_2)}\nonumber\\
&\times\mbox{Tr}\{ \rho_H \left[\langle P_H(t_1) P_H(t_2)\rangle 
-\langle P \rangle^2 \right] \},
\label{H-rep}
\end{align}
where the density matrix $\rho_H$ is time independent in the Heisenberg 
representation. In the Schrodinger representation, the density matrix has 
time dependence.
We now change from the Heisenberg representation to the Schrodinger representation
(denoted by the subscript $S$). Recall that an operator $A_H(t)$ in the
Heisenberg representation can be expressed in the Schrodinger representation by 
$A_H(t)=U^{\dagger}(t,t_0)A_S(t)U(t,t_0)$, where $U(t,t_0)$ is the unitary time 
evolution operator. Hence 
\begin{eqnarray}
\mbox{Tr}\{\rho_H P_H(t_1) P_H(t_2)\} &=&
\mbox{Tr}\{\rho_S(t_1) P_S(t_1) U(t_1,t_2) \nonumber\\
&& P_S(t_2)U(t_2,t_1)\} \nonumber\\
&\equiv& F(t_1,t_2).
\end{eqnarray}
To simplify the notation we have temporarily omitted the symbol
$\langle ...\rangle$ denoting the time average.
The spectral density of polarization fluctuations in the Schrodinger
representation becomes:
\begin{equation}
S_P(\omega) = 
2\int_{-\infty}^{+\infty}\mbox{d}(t_1-t_2)\mbox{e}^{i\omega (t_1-t_2)}
F(t_1,t_2) - 2\langle P\rangle^{2}\delta(\omega),
\label{S-rep}
\end{equation}
where we assume that the system is in a stationary state so that
the function $F(t_1,t_2)$ depends on $(t_1-t_2)$. 
It can be expressed as 
\begin{align}
F(t_1,t_2) &= \sum_{m,n,p}\langle m(t_1)|\rho(t_1)| n(t_1)\rangle 
\langle n(t_1)| P(t_1)| p(t_1)\rangle \nonumber\\
&\times \langle p(t_1)|U(t_1,t_2) P(t_2) U(t_2,t_1)
| m(t_1)\rangle \nonumber\\    
&= \sum_{m,n,p}\langle m(t_1)|\rho(t_1)| n(t_1)\rangle
\langle n(t_1)| P(t_1)| p(t_1)\rangle \nonumber\\
&\times \langle p(t_2)| P(t_2)|m(t_2)\rangle,
\label{eq:F12}
\end{align}
where $\rho(t)$ is the density matrix in the Schrodinger representation,
and $m$, $n$, and $p$ denote eigenstates of $H_0$. 

As we mentioned in the introduction, we are considering
the density matrix $\rho$ of a single TLS that contains the time
dependence of the external driving field. The random dipole fluctuations
are contained in $P_H(t)$. Let $\alpha$ stand for
$m$, $n$, or $p$. Then
$|\alpha(t)\rangle=\mbox{exp}(-i E_{\alpha}t/\hbar)|\alpha \rangle$ 
and $H_0|\alpha\rangle = E_{\alpha}|\alpha\rangle$. 
We now switch from $|n(t)\rangle$, $|m(t)\rangle$, and $|p(t)\rangle$
to the $|+\rangle$ and $|-\rangle$ eigenstates of a TLS to obtain:
\begin{align}
F(t_1,t_2) &= [\rho(t_1) P (t_1)]_{++} P_{++}(t_2) \nonumber\\
&+  [\rho(t_1) P (t_1)]_{--} P_{--}(t_2) \nonumber\\
&+ \mbox{e}^{-i \omega_0 (t_1-t_2)} 
[\rho(t_1) P(t_1)]_{-+} P_{+-}(t_2)\nonumber\\
&+ \mbox{e}^{+i \omega_0 (t_1-t_2)}
[\rho(t_1) P(t_1)]_{+-} P_{-+}(t_2),
\end{align}
where $ P_{\alpha \alpha^{\prime}}(t)$
denotes the $\alpha \alpha ^{\prime}$th element of the $P(t)$ matrix,
$[\rho(t) P(t)]_{\alpha \alpha^{\prime}}$ represents the 
$\alpha \alpha ^{\prime}$th element of the $\rho(t) P(t)$ matrix, 
and $\omega_0\equiv E/\hbar$. We will see in Section \ref{B-R} that $\rho_{+-}$ 
and $\rho_{-+}$ are first order in the small parameter
$\eta = {\bf p}\cdot\bm{\xi}_{ac}/E$. 
$\eta \ll 1$ for both small and large values of $I/I_c$. 
So we will neglect terms with $\rho_{+-}$ and $\rho_{-+}$.
This leads to an approximate expression for $F(t_1,t_2)$:
\begin{align}
\label{noise_integrand}
F(t_1,t_2) &\approx \rho_{++} P_{++}(t_1) P_{++}(t_2) \nonumber\\
&+  \rho_{--} P_{--}(t_1) P_{--}(t_2)  \nonumber\\
&+ \mbox{e}^{-i \omega_0(t_1-t_2)}\rho_{--}
P_{-+}(t_1) P_{+-}(t_2)\nonumber\\
&+ \mbox{e}^{+i \omega_0 (t_1-t_2)}\rho_{++}
P_{+-}(t_1) P_{-+}(t_2).
\end{align}

Let $P_{||}(t)$ be the projection along the ac external field of the 
polarization operator associated with the dipole moment ${p}$ of 
a two level system. $P_{||}(t)$ has stochastic fluctuations due 
to the fact that the electric dipole moment of the two level system 
randomly changes its orientation angle $\theta (t)$ with respect to the applied 
electric field. Hence, in the TLS energy eigenbasis, we can write
\begin{align}
P_{||}(t) &= -\frac{p\cos(\theta (t))}{V_o} 
\biggr (\frac{\Delta_0}{E}\sigma_x + 
\frac{\Delta}{E}\sigma_z \biggr )\nonumber\\
&\equiv P_0(t)\biggr (\frac{\Delta_0}{E}\sigma_x +
\frac{\Delta}{E}\sigma_z \biggr),
\end{align}
where $P_0(t)\equiv-p\cos(\theta(t))/V_o$ and $V_o$ is volume. 

Substituting $P_{||}$ for $P$ in Eq.~(\ref{S-rep}) and
using Eq.~(\ref{noise_integrand}), we obtain 
\begin{align}
S_{P_{||}}(\omega) &= 
2\int_{-\infty}^{+\infty}\mbox{d}(t_1-t_2)\mbox{e}^{i\omega (t_1-t_2)}
\langle P_0(t_1)P_0(t_2) \rangle \nonumber\\
&\times \biggr\{ \biggr(\frac{\Delta}{E}\biggr)^2 
+ \biggr(\frac{\Delta_0}{E}\biggr)^2 \bigr[ \mbox{e}^{-i\omega_0(t_1-t_2)}\rho_{--}\nonumber\\
&+ \mbox{e}^{i\omega_0(t_1-t_2)} \rho_{++} \bigr]\biggr \}
-2\langle P\rangle^2 \delta(\omega),
\end{align}
Since for stationary processes the correlator $\langle P_0(t_1)P_0(t_2)\rangle$ 
is a function of $(t_1-t_2)$, we can define 
$S_{P_0}(\omega)\equiv 2\int_{-\infty}^{+\infty}
\mbox{d}(t_1-t_2)\mbox{e}^{i\omega (t_1-t_2)}
\langle P_0(t_1)P_0(t_2) \rangle $. Then we have 
\begin{align}
S_{P_{||}}(\omega) &= \biggr(\frac{\Delta}{E}\biggr)^2 
S_{P_0}(\omega) \nonumber\\
&+ \biggr(\frac{\Delta_0}{E}\biggr)^2 
\bigr[\rho_{--}S_{P_0}(\omega - \omega_0) + \rho_{++}S_{P_0}(\omega + \omega_0)\bigr]\nonumber\\
&-2\langle P\rangle^2 \delta(\omega).
\label{eq:noise1TLS}
\end{align}
This is a general formula for the spectral density of the polarization fluctuations 
assuming that the fluctuations in the orientations 
of the electric dipole moments of TLS are a stationary process. 
The last term ensures that $S_{P_{||}}(\omega=0)=0$. 
The first term, which is proportional 
to $S_{P_{0}}(\omega)$, is the $relaxation$ (REL) contribution.
It is associated with the TLS pseudospin $\sigma_z$ whose expectation value
is proportional to the population difference between the two levels of the TLS. 
The relaxation contribution to phonon or photon attenuation is due to the
modulation of the TLS energy splitting $E$ by the incident photons which 
have energy $\hbar\omega \ll E$. This modulation causes the population 
of the TLS energy levels to readjust which consumes energy and leads to 
attenuation of the incident electromagnetic flux. 
Because $\rho_{++}+\rho_{--}=1$, the relaxation term has no dependence on the density
matrix, and so will not be affected by saturation effects. Since, as we will
see in section \ref{Results}, this term dominates at 
low frequencies, this implies that the low frequency noise will not be affected
by TLS saturation.

The middle two terms in Eq.~(\ref{eq:noise1TLS}) are proportional to 
$S_{P_0}(\omega \pm \omega_0)$ and are {\it resonance} (RES) contributions. 
The resonance terms are associated with the $x$ and $y$ 
components of the TLS pseudospin that describe transitions between energy levels. 
They describe the resonant absorption by TLS 
of photons or phonons with $\hbar \omega=E$. 
We will see in section \ref{Results} that 
the resonance contributions are dominant at high frequencies. 

To obtain the charge noise $S_Q(\omega)$ from the polarization
noise, we make use of the following formulas.
In a polarized medium, the induced (bound) charge is 
\begin{equation}
Q=\int \bm{P}\cdot d\bm{A}
\end{equation} 
where $\bm{P}$ is the electric polarization. We choose $P_{z}=P_{||}$ and 
$d\bm{A} \| \hat{z}$ since $Q\sim |p_z|=|p\cos\theta|$.
Then 
\begin{eqnarray}
S_Q(\omega)&=&A^2S_{P_{z}}(\omega)\\
&=& A^2S_{P_{||}}(\omega),
\label{eq:charge_noise}
\end{eqnarray}
where $A$ is the area of a plate of a parallel plate capacitor with capacitance
$C=\varepsilon^{\prime} A/L$, and $\varepsilon^{\prime}$ is the 
real part of the dielectric permittivity.

In this section we have derived expressions for the polarization and charge noise
of a single TLS in terms of the density matrix. In the next section we will solve
the Bloch--Redfield equations for the time dependent density matrix of a TLS
subjected to an external ac driving field.

\section{The Bloch-Redfield Equations}
\label{B-R}

From Eqs.~(\ref{S-rep}) and (\ref{eq:F12}), we see that we need the time
dependent density matrix to calculate the polarization noise spectrum.
In this section, we solve the Bloch-Redfield equations to find the time
evolution of the density matrix of a single TLS subject to an external
ac electric field. These equations combine the equation of motion of the 
density matrix with time-dependent perturbation theory, taking into account 
the relaxation and dephasing of TLS.

We follow Slichter\cite{Slichter} and write the following 
set of linear differential equations for the density matrix elements 
$\rho_{\alpha\alpha^{\prime}}(t)$:
\begin{align}
\label{BR1}
\frac{\mbox{d} \rho_{\alpha\alpha^{\prime}}}{\mbox{d}t} &=
\frac{\mbox{i}}{\hbar}\langle\alpha|[\rho, H_0]|\alpha^{\prime} \rangle \nonumber\\
&+ \frac{\mbox{i}}{\hbar}\langle\alpha|[\rho, H_1(t)]|\alpha^{\prime} \rangle \nonumber\\
&+ \sum_{\beta,\beta^{\prime}} R_{\alpha \alpha^{\prime},\beta \beta^{\prime}}
[\rho_{\beta\beta^{\prime}} - \rho^{eq}_{\beta\beta^{\prime}}(T)],
\end{align}
where $\alpha$ and $\beta$ can be either $+$ or $-$, corresponding to
the energy eigenstates of the TLS, and 
$R_{\alpha \alpha^{\prime},\beta \beta^{\prime}}$ are the Bloch-Redfield
tensor components which are constant in time. They are related to the
longitudinal and transverse relaxation times, $T_1$ and $T_2$. 
In Eq.~(\ref{BR1}), $H_0$ and $H_1(t)$ are Hamiltonians given by
Eqs.~(\ref{diagH0}) and (\ref{eq:diagH1}), respectively. 
The thermal equilibrium value of the density matrix is denoted by 
$\rho^{eq}(T)$. 

In thermal equilibrium only the diagonal elements of the density matrix 
are nonzero, and are given by $\rho^{eq}_{--}(T)=\exp(+E/2k_BT)/Z$
and $\rho^{eq}_{++}(T)=\exp(-E/2k_BT)/Z$, where the partition 
function $Z=\left[\exp(-E/2k_BT)+\exp(+E/2k_BT)\right]$.
Here we are using the fact that in thermal equilibrium, the density 
matrix can be represented as
\begin{equation}
\hat{\rho}_{eq}(T)=\frac{1}{Z(T)}e^{-\hat{H}_{0}/k_BT}
\label{eq:eqdensitymatrix}
\end{equation}
One may wonder whether instead we should use 
\begin{equation}
\hat{\rho}(T,t)=\frac{1}{Z(T,t)}\exp
\left[-\left(\hat{H}_{0}+\hat{H}_{1}(t)\right)/k_BT\right]
\label{eq:acdensitymatrix}
\end{equation}
since the ac field changes the TLS energy splitting. 
Slichter has discussed this issue in his book.\cite{Slichter}  
Eq.~(\ref{eq:eqdensitymatrix}) is appropriate if the TLS are too slow to respond
to the external ac field, but Eq.~(\ref{eq:acdensitymatrix}) should be used if the
external field varies much more slowly than the response time of the TLS.
In the latter case, the external field looks like a static field to the TLS.
For the cases of interest, typical experimental ac external 
fields operate at several GHz while the response or relaxation time of TLS is $T_1$ 
which, as we said earlier, typically varies between 10$^{-9}$ s 
and 10$^{4}$ s. So it is reasonable to use Eq.~(\ref{eq:eqdensitymatrix}).

Eq.~(\ref{BR1}) can be written in the form:
\begin{align}
\label{BR2}
\frac{\mbox{d} \rho_{\alpha\alpha^{\prime}}}{\mbox{d}t}
&= \frac{\mbox{i}}{\hbar}(E_{\alpha^{\prime}}-E_{\alpha})\rho_{\alpha \alpha^{\prime}} +
\frac{\mbox{i}}{\hbar}\sum_{\alpha^{\prime\prime}}[\rho_{\alpha\alpha^{\prime\prime}}
\langle\alpha^{\prime\prime}|H_1(t)|\alpha^{\prime} \rangle \nonumber\\
&- \langle\alpha|H_1(t)|\alpha^{\prime\prime} \rangle
\rho_{\alpha^{\prime\prime}\alpha^{\prime}}]\nonumber\\
&+ \sum_{\beta,\beta^{\prime}} R_{\alpha \alpha^{\prime},\beta \beta^{\prime}}
[\rho_{\beta\beta^{\prime}} - \rho^{eq}_{\beta\beta^{\prime}}(T)].
\end{align}
Next we use the fact that in the relaxation terms
$R_{\alpha \alpha^{\prime},\beta \beta^{\prime}}$, the only important terms
correspond to\cite{Slichter,Redfield1957} 
$\alpha-\alpha^{\prime}=\beta-\beta^{\prime}$.
In addition, the Bloch-Redfield tensor is symmetric, so we have the following 
relations for the dominant components:
\begin{align}
\label{tau1}
R_{--,++} &= R_{++,--}\equiv\frac{1}{T_1}\\
\label{tau2}
R_{-+,-+} &= R_{+-,+-}\equiv-\frac{1}{T_2}.
\end{align}
The longitudinal relaxation time $T_1$ is given by Eq.~(\ref{eq:T1inv}). For
the transverse relaxation time $T_2$, we will use the experimental value
\cite{Bernard1979}
\begin{equation}
T_{2}\approx \frac{8\times 10^{-7}}{T}\;\;{\rm sec},
\end{equation}
where $T$ is in Kelvin. From relations (\ref{tau1}) and (\ref{tau2}), 
the set of linear differential equations (\ref{BR2}) becomes
\begin{align}
\frac{\mbox{d} \rho_{--}}{\mbox{d}t}
& = \frac{\mbox{i}}{\hbar}
( \rho_{-+}[H_1(t)]_{+-} - [H_1(t)]_{-+}\rho_{+-} ) \nonumber\\
& + \frac{1}{T_1}( \rho_{++} - \rho^{eq}_{++}(T) ),\\
\frac{\mbox{d} \rho_{++}}{\mbox{d}t}
& = \frac{\mbox{i}}{\hbar}
( \rho_{+-}[H_1(t)]_{-+} - [H_1(t)]_{+-}\rho_{-+} ) \nonumber\\
& + \frac{1}{T_1}( \rho_{--} - \rho^{eq}_{--}(T) ),\\
\frac{\mbox{d} \rho_{-+}}{\mbox{d}t}
& = \frac{\mbox{i}}{\hbar}(E_{+}-E_{-})\rho_{-+} \nonumber\\
& + \frac{\mbox{i}}{\hbar}( \rho_{--}[H_1(t)]_{-+} - [H_1(t)]_{--}\rho_{-+} \nonumber\\
& + \rho_{-+}[H_1(t)]_{++} - [H_1(t)]_{-+}\rho_{++} ) -\frac{1}{T_2}\rho_{-+},\\
\frac{\mbox{d} \rho_{+-}}{\mbox{d}t}
& = \frac{\mbox{d} \rho_{-+}^{\star}}{\mbox{d}t},
\end{align}
where $[H_1(t)]_{\alpha,\beta}=\langle \alpha|H_1(t)| \beta\rangle$.
Using Eq.~(\ref{eq:diagH1}), we can write the first two equations 
for the diagonal elements as:
\begin{align}
\frac{\mbox{d} \rho_{--}}{\mbox{d}t} &= - \frac{\mbox{d} \rho_{++}}{\mbox{d}t} =
\frac{\mbox{i}}{\hbar}(-\eta \Delta_0 \mbox{cos} \Omega t)
~(\rho_{-+} - \rho_{+-})\nonumber\\
&+ \frac{1}{T_1}[\rho_{++} - \rho_{--} -(\rho^{eq}_{++}(T)-\rho^{eq}_{--}(T))],
\end{align}
while the equation for the off--diagonal element $\rho_{-+}$ is:
\begin{align}
\frac{\mbox{d} \rho_{-+}}{\mbox{d}t} &=
\frac{\mbox{i}}{\hbar}(E_{+}-E_{-})\rho_{-+} + \frac{\mbox{i}}{\hbar} 
(\rho_{--}-\rho_{++})(-\eta \Delta_0 \mbox{cos}\Omega t)\nonumber\\
&+ \frac{\mbox{i}}{\hbar}\rho_{-+}(-2 \eta \Delta \mbox{cos}\Omega t) 
- \frac{1}{T_2}\rho_{-+}.
\end{align}

We look for a steady state solution of the form
\begin{align}
\rho_{--} &= r_{--}, ~~~ \rho_{++} = r_{++}\\
\rho_{-+}(t) &= r_{-+} \mbox{e}^{\mbox{i}\Omega t},
~~~\rho_{+-}(t) = \rho_{-+}^{\star}(t),
\end{align}
where $r_{\alpha\alpha^{\prime}}$ are complex constants. We find that 
in steady state the density matrix elements are given by the following 
expressions:
\begin{align}
\rho_{--} &= \frac{1}{2} + \frac{\rho^{eq}_{--}(T)-1/2}{1 + g(\Omega,\omega_0,T_2)\times (I/I_c)}
\label{rhomm}\\
\rho_{++} &= \frac{1}{2} + \frac{\rho^{eq}_{++}(T)-1/2}{1 + g(\Omega,\omega_0,T_2)\times (I/I_c)}
\label{rhopp}\\
\rho_{-+}(t) &= - \frac{\mbox{i}T_2\eta \Delta_0}{2\hbar}\frac{1}
{[1+\mbox{i}T_2(\Omega-\omega_0)]}\nonumber\\
&\times \frac{\rho^{eq}_{--}(T)-\rho^{eq}_{++}(T)}{1 + g(\Omega,\omega_0,T_2)\times (I/I_c)}
~\mbox{exp}(\mbox{i}\Omega t)\label{rhomp}\\
\rho_{+-}(t) &= \rho_{-+}^{\star}(t),
\label{rhopm}
\end{align}
where $\eta= {\bf p} \cdot \bm{\xi}_{ac}/E$,
$I/I_c=T_1 T_2 (\eta \Delta_0/\hbar)^2/2$, and
$g(\Omega,\omega_0,T_2)=1/[1+(\Omega-\omega_0)^2T_2^2]$.
For a dipole moment $p=3.7$ D, a large electric field $\xi_{ac}=3\times 10^3$
V/m, and TLS energy splittings of the order of 10 GHz,
the dimensionless factor $\eta\approx 0.005$, and it decreases 
to a value of $5\times 10^{-8}$ when the amplitude of the applied 
electric field is $\xi_{ac}=0.03$ V/m. $g(\Omega,\omega_0,T_2)$ is 
approximately equal to $1$ when the ac driving frequency is resonant with 
the TLS energy splitting, i.e., $\Omega\approx \omega_0$. 

Notice that the off-diagonal elements of the density matrix 
are first order in $\eta \ll 1$. They are oscillatory and small, as shown 
by the following numerical estimate. For large electric fields
($\xi_{ac}=3\times 10^3$ V/m), $\eta\approx 0.005$, $T_2=8$ $\mu$s at $T=0.1$ K,
\cite{Bernard1979,herve} $\Delta_0/h\approx E/h \approx 10$ GHz, and
$T_1=8\times 10^{-8}$ s,
we obtain $I/I_c=T_1 T_2 (\eta \Delta_0/\hbar)^2/2\approx 10^7\gg 1$, and
$|\rho_{-+}|\approx 10^{-4}$. For $I\ll I_c$, 
we have $|\rho_{-+}|\approx 10^{-2}$. Hence, the amplitude 
of the off-diagonal density matrix elements is very small for both
$I\gg I_c$ and $I\ll I_c$. 

On the other hand, the diagonal elements recover their equilibrium values 
($\rho^{eq}_{--}(T)=\exp(+E/2k_BT)/Z$ and 
$\rho^{eq}_{++}(T)=\exp(-E/2k_BT)/Z$) in the
limit of low electromagnetic fields $I\ll I_c$. For large electric fields 
$I\gg I_c$, they approach their steady state values $\rho_{++}=\rho_{--}=1/2$.
As required,
$\mbox{Tr}\{\rho \}= \rho_{--}+\rho_{++}=1$. In addition, the quantum 
expectation value of the $z-$component of the TLS spin is 
\begin{eqnarray}
\langle S_z \rangle &=& \left(\rho_{++}-\rho_{--}\right)\\
&=& -\frac{\tanh(E/(2k_BT))}{[1+(I/I_c)\times g(\Omega,\omega_0,T_2)]}.
\end{eqnarray}

All the elements of the density matrix depend on the 
ratio $I/I_c(\Omega,T)$. We can write approximate expressions for 
$\rho_{--}$ and $\rho_{++}$ corresponding to the populations of
the lower and upper TLS energy levels for both small and large 
electromagnetic fields. 
For $I \ll I_c$ and $g\approx 1$, we expand $\rho_{--}$ and $\rho_{++}$ to 
first order in $I/I_c$ to obtain: 
\begin{align}
\rho_{--} &\approx \rho^{eq}_{--}(T)\bigr( 1- \frac{I}{I_c}g \bigr)
+ \frac{1}{2} \frac{I}{I_c}g\\
\rho_{++} &\approx \rho^{eq}_{++}(T)\bigr( 1- \frac{I}{I_c}g \bigr)
+ \frac{1}{2} \frac{I}{I_c}g.
\end{align}
This means that {\it unsaturated} TLS have $\rho_{++}\approx 0$ and 
$\rho_{--}\approx 1$. Thus the upper level is mostly unoccupied 
while the lower level is almost always occupied.

On the other hand, for $I \gg I_c$, we can expand the steady state solution for 
$\rho_{--}$ and $\rho_{++}$ given in Eqs.~(\ref{rhomm}) and (\ref{rhopp})
to first order in $(I/I_c)^{-1}$. The result is
\begin{align}
\rho_{--} &\approx \frac{1}{2} + \bigr[\rho^{eq}_{--}(T) - \frac{1}{2}\bigr]
\bigr( \frac{I}{I_c}g \bigr)^{-1}\\
\rho_{++} &\approx \frac{1}{2} - \bigr[\rho^{eq}_{--}(T) - \frac{1}{2}\bigr]
\bigr( \frac{I}{I_c}g \bigr)^{-1}.
\end{align}
Hence, for {\it saturated} TLS the lower and upper levels are almost equally 
populated, i.e., $\rho_{++}\approx \rho_{--}\approx 1/2$. 
This is to be expected since the TLS are constantly being excited by the
ac electric field and de-excited by spontaneous and stimulated emission.
Once the TLS are saturated, the populations of the lower and upper levels 
will have small deviations from their steady state values. We will look at 
the saturation effect in more detail in Section \ref{Results} where we 
plot the noise spectrum of a single fluctuator versus $I/I_c$. 
This noise spectrum depends on the density matrix elements $\rho_{++}$ 
and $\rho_{--}$. As one goes from the unsaturated regime to the saturated regime, 
the amplitude of $\rho_{--}$ decreases by a factor of two. 
From Eq.~(\ref{eq:noise1TLS}) the polarization noise of a 
single TLS depends linearly on $\rho_{--}$. 
Because $\rho_{--}$ only decreases 
by a factor of 2 when the TLS are saturated, we will see that the 
saturation of TLS will not play an important role in the polarization
and charge noise spectra. 

\section{Results}
\label{Results}
In this section we begin by studying the polarization noise spectrum of a single TLS 
fluctuator as a function of frequency and the electric field intensity ratio ($I/I_c$). 
We then obtain the total polarization noise by averaging over 
the distribution of independent two level systems. From this we get the
charge noise that we will analyze at both low and high frequencies as a function
of temperature and electric field intensity.

\subsection{Polarization Noise of One TLS Fluctuator}
Now that we have the matrix elements of the steady state density matrix 
in Eqs.~(\ref{rhomm})--(\ref{rhopm}), we can use them to evaluate the 
expression for polarization noise found in Eq.~(\ref{eq:noise1TLS}).
In order to make further progress in evaluating Eq.~(\ref{eq:noise1TLS}), 
we need to know the polarization noise spectrum $S_{P_0}(\omega)$ of a 
fluctuating TLS. So we assume that a single TLS fluctuates
randomly in time. Its electric dipole moment fluctuates in orientation
by making 180$^o$ flips between $\theta=\theta_0$ and
($\theta=\theta_0+180^o$), resulting
in a random telegraph signal (RTS) in the polarization 
$P_0(t)=\pm p\cos(\theta_{0})/V$ along the external
field. $S_{P_0}(\omega)$ dominates at low frequencies when the system is
in thermal equilibrium. So for $S_{P_0}(\omega)$, we will use a Lorentzian
noise spectrum given by\cite{Machlup,Kogan96} 
\begin{equation}
S_{P_0}(\omega)=4w_1w_2\frac{\langle \delta P_0^2\rangle \tau}
{(1 + \omega^2\tau^2)},
\end{equation} 
where $\tau$ is the characteristic relaxation time of the fluctuator,
and $w_1$ ($w_2$) is the probability of being in the lower
(upper) state of the TLS. Since the ratio of the probabilities
of being in the upper versus lower state is $w_2/w_1=\exp(-E/k_BT)$,
and $w_1+w_2=1$, the product
\begin{equation}
4w_1w_2=\frac{1}{\cosh^{2}(E/2k_BT)} 
\end{equation}
So in Eq.~(\ref{eq:noise1TLS}), we replace $S_{P_0}(\omega)$ by the 
RTS noise spectrum with a relaxation time $\tau=T_1$ since this term is 
associated with $\sigma_z$, the longitudinal component of the TLS pseudospin. 

At high frequencies, $S_{P_0}(\omega\pm\omega_0)$ dominates 
and is associated with resonant processes. At high intensities when
the driving frequency $\Omega$ is close to the energy splitting $\omega_0$,
saturation occurs, and the system is not in thermal equilibrium. So we will use
\begin{equation}
S_{P_0}(\omega\pm\omega_0)=\frac{\langle \delta P_0^2\rangle \tau}
{(1 + (\omega\pm\omega_0)^2\tau^2)},
\end{equation}
Since these terms are associated with the transverse component $\sigma_x$
of the TLS pseudospin, we will set $\tau$ equal to the transverse relaxation 
time $T_2$.

Putting this all together, we arrive at the following expression for 
the spectral density of polarization fluctuations of a single TLS:
\begin{align}
\frac{S_{P_{||}}(\omega)}{\langle\delta P_0^2\rangle} &= 
\biggr(\frac{\Delta}{E}\biggr)^2 \bigr[\frac{1}{\cosh^2(E/2k_BT)}] 
\frac{T_1}{1+\omega^2T_1^2} \nonumber\\
& + \biggr(\frac{\Delta_0}{E}\biggr)^2 \biggr[(\rho_{--})\frac{T_2}
{1+(\omega-E/\hbar)^2T_2^2} \nonumber\\
& +(\rho_{++})\frac{T_2}{1+(\omega+E/\hbar)^2T_2^2}\biggr]\label{noise1RTS}
+{\rm dc\;term}
\end{align}
The first term is the relaxation contribution and the middle two terms are
the resonance contribution. The dc term ensures that $S_{P_{||}}(\omega=0)=0$.

\begin{figure}
\includegraphics[height = 5.5cm, width = 7.5cm]{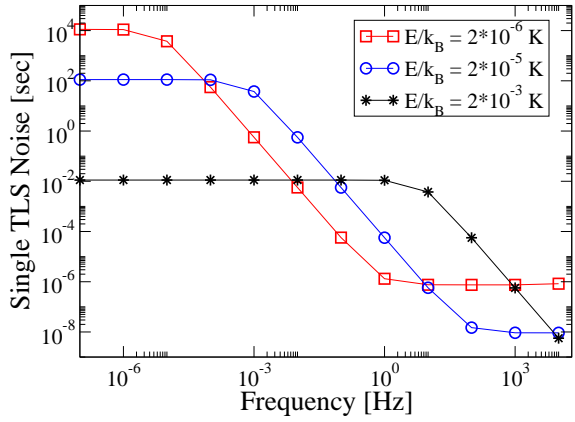}
\caption{\label{RTSlowFreq} (Color online)
Log-log plot of the low frequency polarization 
noise of a single two level system 
given by Eq.~(\ref{noise1RTS}) versus frequency for three different 
values of the TLS energy splitting as shown in the legend. For all
3 cases, $T=0.1$ K, $I/I_c=10^{-4}$, $\Omega=E/\hbar$, and 
$\Delta=\Delta_0=E/\sqrt{2}$.\\
\vspace{0.5cm}}
\end{figure}

In Fig.~\ref{RTSlowFreq}, we show the spectral density 
$S_{P_{||}}(\omega)/\langle\delta P_0^2\rangle$ of polarization fluctuations
of a single TLS at low frequencies. We consider three values 
of the TLS energy splitting as shown in the legend. 
In the low frequency range, the noise
spectrum is dominated by the relaxation contribution which is a Lorentzian.
Thus the noise spectrum is flat for $\omega \ll 1/T_1$. As the frequency 
increases, it rolls over at $\omega \sim 1/(T_{1})$, and goes 
as $1/\omega^2$ for $\omega > 1/T_1$. At very high frequencies
$\left(\omega > |E/\hbar|\right)$, 
it saturates to a constant (white noise) due to the resonance terms.
In addition, we find that the low frequency noise is independent 
of electric field intensity ratio $I/I_c$ and the angular frequency $\Omega$ 
of the ac driving field. 

\begin{figure}
\includegraphics[height = 5.5cm, width = 7.5cm]{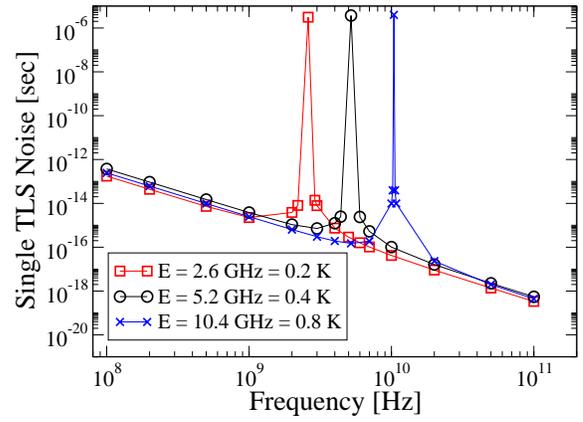}
\caption{\label{RTS_RES} (Color online)
Log-log plot of the high frequency polarization 
noise of a single two level system 
given by Eq.~(\ref{noise1RTS}) versus frequency for three different 
TLS energy splittings. The peaks appears when the resonance condition 
$\omega=E/\hbar$ is satisfied. In all three cases, $T=0.1$ K, $I/I_c=10^{-4}$, 
$\Omega=E/\hbar$ and $\Delta=\Delta_0=E/\sqrt{2}$. No noticeable difference 
was obtained for $I/I_c=10^{7}\gg 1$ (not shown).}
\end{figure}

The contribution of the resonance terms to the noise
spectrum is negligible at low frequencies ($\omega\ll E/\hbar$) as expected
from simple numerical estimates. However these resonance terms become
important at high frequencies as shown in Fig.~\ref{RTS_RES}.
The peaks appear when the resonance condition $\omega=\omega_0= E/\hbar$ is 
satisfied. The plot in Fig.~\ref{RTS_RES} was obtained for the low value 
of the ratio $I/I_c=10^{-4}$. No noticeable differences were obtained 
when the ratio was increased to values as high as $10^7$ (not shown). 

\begin{figure}
\includegraphics[height = 5.5cm, width = 7.5cm]{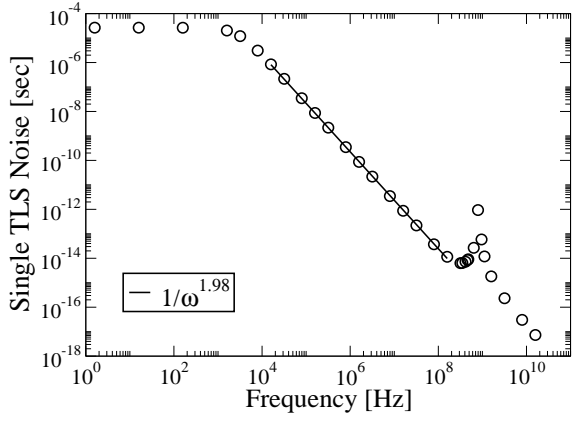}
\caption{Log-log plot of the polarization noise of a single two level system
given by Eq.~(\ref{noise1RTS}) versus frequency for a broad range of 
frequencies. $E$ = 5.2 GHz = 0.4 K, $I/I_c=10^{-4}$,
$\Omega=E/\hbar$, and $\Delta=\Delta_0=E/\sqrt{2}$.}
\label{SingleTLSNoise_Hi_LowFreq}
\end{figure}

Figure \ref{SingleTLSNoise_Hi_LowFreq} shows the polarization noise power
of a single two level system over a broad range of frequencies that covers
both the resonance and relaxation contributions. At low frequencies there
is a plateau. Between 100 kHz and 1 GHz, the noise spectrum decreases as
$1/\omega^{2}$ due to the Lorentzian associated with the relaxation
contribution. At higher frequencies there is a resonance peak at
$\omega=E/\hbar$.

\begin{figure}
\includegraphics[height = 5.5cm, width = 8cm]{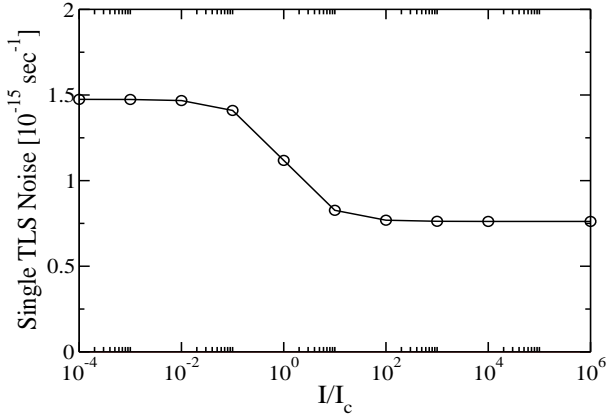}
\caption{Log-linear plot of the high frequency polarization 
noise for a single two level system given by 
Eq.~(\ref{noise1RTS}) versus $I/I_c$. 
The amplitude of the noise decreases by a factor of 
two as saturation is achieved for $I/I_c>1$. $\omega/2\pi=9$ GHz,
$E/h=\Omega/2\pi=10$ GHz, $\Delta=\Delta_0=E/\sqrt{2}$, and $T=0.1$ K.}
\label{Jdep}
\end{figure}

The effect of TLS saturation can be seen at high frequencies in the 
plot of the noise of a single TLS versus the ratio $I/I_c$ as shown in 
Fig.~\ref{Jdep}. We plotted the spectral density of polarization 
fluctuations of an individual TLS given by Eq.~(\ref{noise1RTS}) 
at a fixed high frequency ($\omega/2\pi=9$ GHz) versus $I/I_c$. 
Notice that the noise is constant as long as $I\ll I_c$, then decreases 
when the electric field intensity increases to a value
comparable to the critical intensity, and reaches a value which is half 
the previous one for $I\gg I_c$. This is in agreement with 
our estimates from Section \ref{B-R} where we saw that as $I/I_c$ 
increases, $\rho_{--}$ decreases by a factor of 2 from a value close 
to 1 to a value close to 0.5 when the TLS are saturated.

\subsection{Polarization and Charge Noise of an Ensemble of TLS Fluctuators}

Until now we have analyzed the contribution of a single fluctuator 
to the polarization noise spectrum. We now average the polarization
noise of a single fluctuator over an ensemble of independent fluctuators 
in a volume $V_o$. The distribution function over TLS parameters was given
in Eq.~(\ref{eq:PTLS}) as $P(\Delta,\Delta_0)=P_{TLS}/\Delta_0$. 
Using this, we obtain:
\begin{align}
\frac{S_{P_{||}}(\omega)}{\langle \delta P_0^2\rangle} & = 
V_o \int_{\Delta_{min}}^{\Delta_{max}}\mbox{d} \Delta
\int_{\Delta_{0,min}}^{\Delta_{0,max}}\mbox{d} \Delta_0 P(\Delta,\Delta_0)\nonumber\\ 
&\times\biggr(\frac{\Delta}{E}\biggr)^2 \bigr[\frac{1}{\cosh^2(E/2k_BT)}]
\frac{T_1}{1+\omega^2T_1^2} \nonumber\\
& + \biggr(\frac{\Delta_0}{E}\biggr)^2 \biggr[(\rho_{--})\frac{T_2}
{1+(\omega-E/\hbar)^2T_2^2} \nonumber\\
& +(\rho_{++})\frac{T_2}{1+(\omega+E/\hbar)^2T_2^2}\biggr]
+{\rm dc\;term}\nonumber\\
&\equiv V_o J(\omega; \Omega, T, I/I_c) +{\rm dc\;term},
\label{manyTLS}
\end{align}
where $J(\omega; \Omega, T, I/I_c)$ is the result of integrating over the
distribution of TLS parameters $\Delta$ and $\Delta_0$. 

To obtain the charge noise $S_Q(\omega)$ from the polarization noise,
we use Eq.~(\ref{eq:charge_noise}) with 
\begin{eqnarray}
\langle \delta P_0^2\rangle &=& 
\left\langle\left(\frac{2p\cos\theta_0}{V_o}\right)^2\right\rangle \\
&=& \frac{4p^2}{3V_o^2}.
\end{eqnarray}
We obtain 
\begin{equation}
\frac{S_Q(\omega)}{e^2}=\frac{4}{3}\bigr(\frac{p}{eL}\bigr)^2 V_o 
J(\omega; \Omega, T, I/I_c)+{\rm dc\;term}.
\label{S_Q}
\end{equation}
We can evaluate Eq.~(\ref{S_Q}) numerically using $p=3.7$ D, 
$P_{TLS} = 10^{45}$ (Jm$^3$)$^{-1}$, $L=400$ nm,
$A=40\times 800$ nm$^2$, $V_o=AL$, $\Delta_{min}=0$,
$\Delta_{max}/k_B=\Delta_{0,max}/k_B=4$ K, and 
$\Delta_{0,min}/k_B=2\times 10^{-6}$ K. 
The results follow. We will drop the dc term from now on since it only
affects the zero frequency noise.

\subsubsection{Low Frequency Charge Noise}
\begin{figure}
\includegraphics[height = 5.5cm, width = 8cm]{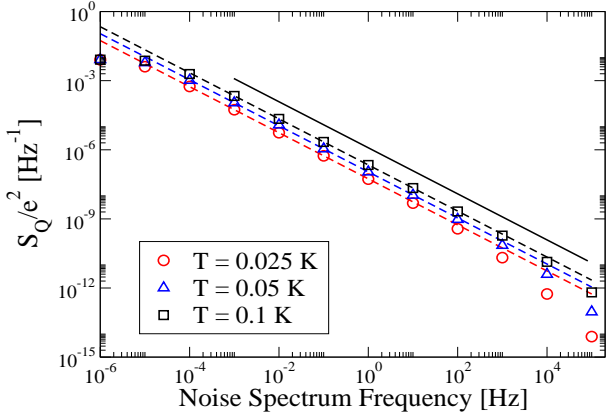}
\caption{\label{fig6} (Color online)
Log-log plot of the low frequency charge noise 
$S_Q/e^2$ averaged over the TLS ensemble
versus frequency for different temperatures. 
The symbols are the results of numerically evaluating Eq.~(\ref{manyTLS}) 
with $I/I_c=10^{-4}$ and $\Omega=10$ GHz. The
dashed lines are the low frequency approximation given by Eq.~(\ref{eq:lowfreq}).
The solid line has a slope of $-1$ corresponding to a perfect $1/f$ spectrum.
Between $10^{-3}$ and $10^4$ Hz, the noise spectrum is very close to $1/f$.
For example, in this frequency range for $T=0.1$ K the noise spectrum goes 
as $1/(f^{1.02})$. 
}
\end{figure}

Evaluating Eq.~(\ref{S_Q}) produces
the normalized low frequency ($\omega\ll |E/\hbar|$)
charge noise spectrum $S_Q/e^2$ 
shown in Fig.~\ref{fig6}. It is flat at very low frequencies 
$\omega T_1 \ll 1$. As the frequency increases,
it rolls over at $\omega\approx (T_{1,max})^{-1}$ and 
decreases as $1/f$ noise between approximately $10^{-3}$ and $10^4$ Hz.  
Here $T_{1,max}$ is the maximum value of $T_1$. For $T=0.1$ K, 
$T_{1,max}\sim 10^{4}$ s.

We can obtain an approximate analytic expression for the low frequency noise from 
Eq.~(\ref{S_Q}) in the following way. By low frequency, we mean that
$\omega\tau_{min}\ll 1$ and $\hbar\omega\ll E$. So we only keep the first
term in $J(\omega; \Omega, T, I/I_c)$ as defined by Eq.~(\ref{manyTLS}):
\begin{eqnarray}
\frac{S_Q(\omega)}{e^2}&\simeq &\frac{4}{3}\bigr(\frac{p}{eL}\bigr)^2 V_o
\int_{\Delta_{min}}^{\Delta_{max}}\mbox{d} \Delta
\int_{\Delta_{0,min}}^{\Delta_{0,max}}\mbox{d} \Delta_0 P(\Delta,\Delta_0)\nonumber\\
&\times &\biggr(\frac{\Delta}{E}\biggr)^2 \bigr[\frac{1}{\cosh^2(E/2k_BT)}]
\frac{T_1}{1+\omega^2T_1^2}
\end{eqnarray}
We change variables to the TLS energy splitting $E$ and the relaxation
time $T_1$. Using
\begin{equation}
\frac{\Delta^{2}}{E^2}=1-\left(\frac{\tau_{min}}{T_1}\right),
\end{equation}
we obtain
\begin{eqnarray}
\frac{S_Q(\omega)}{e^2} &\simeq &\frac{4}{3}\bigr(\frac{p}{eL}\bigr)^2 V_o
\int_{\tau_{min}}^{\infty} dT_1\int_{0}^{E_{max}} dE \; P(E,T_1)\nonumber\\
&\times &\left[1-\frac{\tau_{min}}{T_1}\right]
\frac{1}{\cosh^{2}\left(E/2k_BT\right)} \frac{T_1}{1+\omega^2 T_1^2}
\end{eqnarray}
where $P(E,T_1)$ is given by Eq.~(\ref{eq:PET1}). Using 
$\left(\tau_{min}/T_1\right)\ll 1$ yields
\begin{eqnarray}
\frac{S_Q(\omega)}{e^2} & \simeq & \frac{4}{3}\bigr(\frac{p}{eL}\bigr)^2 
\frac{V_o P_{TLS}}{2}
\int_{0}^{E_{max}} dE \frac{1}{\cosh^2\left(E/2k_BT\right)}\nonumber\\
&\times &\int_{\tau_{min}}^{\infty}\frac{dT_1}{1+\omega^2 T_1^2}\nonumber\\
&\simeq &\frac{2\pi}{3}V_oP_{TLS}\left(\frac{p}{eL}\right)^2\frac{k_BT}{\omega}
\label{eq:lowfreq}
\end{eqnarray}
Thus we obtain $1/f$ noise that goes linearly with temperature. 
As Fig.~\ref{fig6} shows, this expression for the low frequency noise gives 
good agreement with our numerical evaluation of Eq.~(\ref{manyTLS}).
Eq.~(\ref{eq:lowfreq}) also agrees with Kogan\cite{Kogan96}, and Faoro 
and Ioffe\cite{Faoro2006}. To estimate the value of $S_Q/e^2$ from
Eq.~(\ref{eq:lowfreq}),
we use $p=3.7$ D, $P_{TLS} \approx 10^{45}$ (Jm$^3$)$^{-1}$, $L=400$ nm,
$A=40\times 800$ nm$^2$, and $V_o=AL$. At $T=0.1$ K and $f=1$ Hz, we obtain
$S_Q/e^2=2\times10^{-7}$ Hz$^{-1}$, which is comparable to the experimental
value of $4\times 10^{-6}$ Hz$^{-1}$ deduced from current noise.
\cite{Astafiev2006} The magnitude of this noise estimate is also in good
agreement with our numerical result from Eq.~(\ref{S_Q}), i.e.,
$S_Q(f=1$~Hz$)/e^2=2\times 10^{-7}$ Hz$^{-1}$.

We can also obtain this $1/f$ noise result with the following simple 
calculation.  Consider an electric dipole moment ${\bf p}$ in a parallel 
plate capacitor at an angle $\theta_0$
with respect to the $z-$axis which is perpendicular to the electrodes
that are a distance $L$ apart. Assume that the electrodes are at the
same voltage.
When the dipole flips by 180$^{o}$, the induced charges on the 
superconducting electrodes change from $\mp p \cos\theta_0/L$ to 
$\pm p \cos\theta_0/L$. Let $\delta Q$ denote the magnitude of the charge 
fluctuations. Then $\delta Q=|2 p \cos\theta_0/L|$. Hence the charge 
of the Josephson junction 
capacitor produces a simple two state random telegraph signal which switches 
with a transition rate $\tau^{-1}$ given by the sum of 
the rates of transitions up and down.\cite{Kogan96} The charge noise spectral 
density is \cite{Kogan96}
\begin{align}
S_Q^{(i)}(\omega)&=2\int_{-\infty}^{\infty} 
{\mbox{d}(t_1-t_2)\mbox{e}^{i\omega(t_1-t_2)} 
\langle \delta Q(t_1) \delta Q(t_2)\rangle }\nonumber\\
&=\langle(\delta Q)\rangle^2\frac{4 w_1 w_2 \tau}{1+\omega^2\tau^2}~~,
\end{align} 
where the superscript $i$ refers to the $i$th TLS in the substrate or 
tunnel barrier, and $w_1$ ($w_2$) is the probability of being in the lower 
(upper) state of the TLS. In order to average over TLS, we recall 
from Eq.~(\ref{eq:PET1}) that the distribution function of TLS parameters 
can be written in terms of the energy and relaxation times:\cite{Phillips}  
\begin{equation}
P(E,\tau)=\frac{P_{TLS}}{(2\tau\sqrt{1-\tau_{min}/\tau})}.
\end{equation} 
At low frequencies $\omega\tau_{min}\ll 1$, 
the main contribution to the spectral density comes from slowly relaxing
TLS for which $P(E,\tau)\simeq P_{TLS}/(2\tau)$. Therefore, the charge 
noise of an ensemble of independent fluctuators is 
\begin{align}
S_Q(\omega)& \simeq V_o\int_{\tau_{min}}^{\infty} \mbox{d}\tau 
\int_{0}^{E_{max}}\mbox{d}E~\frac{P_{TLS}}{2\tau}\nonumber\\
&\times \frac{\langle \delta Q^2\rangle}{\cosh^2(E/2k_B T)}
\frac{\tau}{1+\omega^2\tau^2} ~~.
\end{align}
$\langle \delta Q^2\rangle$ is the square of the amplitude of charge 
fluctuations averaged over TLS dipole orientations. We assume that 
$\langle \delta Q^2\rangle$ is independent of $E$ and $\tau$. 
At low temperatures ($k_BT \ll E_{max}$) and low frequencies
($\omega\tau_{min}\ll 1$), we recover Eq.~(\ref{eq:lowfreq}) which
describes $1/f$ charge noise that goes linearly with temperature.

Still another way to obtain low frequency $1/f$ noise is the following.
At low frequencies the system is in thermal equilibrium, and we can use 
Eq.~(\ref{eq:S_Q_Tan_delta}) from Section \ref{FDT}: 
\begin{equation}
\frac{S_Q(\omega)}{e^2}=\frac{2k_BT}{e^2/2C}\frac{\tan \delta}{\omega}.
\label{eq:SQtanDeltaLF}
\end{equation}
The TLS contribution to the dielectric loss tangent
$\tan\delta$ was calculated by previous workers 
\cite{Phillips,Classen1994,Arnold1976}
who considered fluctuating TLS with electric dipole moments. They found
\begin{equation}
\tan\delta=\frac{\pi p^2 P_{TLS}}{6\varepsilon^{\prime}}
\label{eq:tanDelta}
\end{equation}
where $\varepsilon^{\prime}$ is the real part of the dielectric 
permittivity and $P_{TLS}$ the constant TLS density of states.  
By plugging Eq.~(\ref{eq:tanDelta}) into Eq.~(\ref{eq:SQtanDeltaLF}) 
and using Eq.~(\ref{eq:Cap}) and $V_o=AL$, we recover Eq.~(\ref{eq:lowfreq}).

\subsubsection{High Frequency Charge Noise}
\begin{figure}
\includegraphics[height = 5.5cm, width = 8cm]{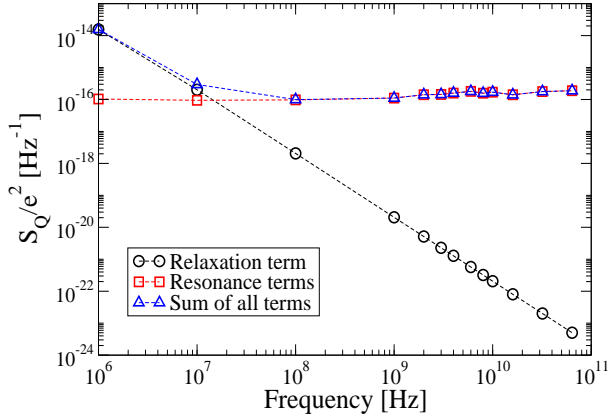}
\caption{\label{fig7} (Color online) 
Log-log plot of the high frequency charge noise 
$S_Q/e^2$ averaged over the TLS ensemble versus frequency. The contributions 
of the relaxation term 
and the resonance terms are plotted separately. 
The noise spectrum becomes white noise at frequencies 
$\stackrel{>}{\sim} 10$ MHz. The results were obtained using $T=0.1$ K, 
$\Omega=10$ GHz, and $I/I_c=10^7$.}
\end{figure}

Evaluating Eq.~(\ref{S_Q}) numerically at high frequencies yields
the normalized charge noise spectrum shown in Fig.~\ref{fig7}. 
We have plotted the contributions coming from the relaxation term and the two 
resonance terms separately. As we described in
the previous subsection, the relaxation term dominates at low frequencies and
gives $1/f$ noise. In the high frequency range shown in Fig.~\ref{fig7},
the relaxation term produces $1/f^2$ noise. On the other hand, the 
resonance terms produce white (flat) noise that dominates at high 
frequencies. The two curves cross at approximately 10 MHz. 

\begin{figure}
\vspace{0.5cm}
\includegraphics[height = 5.5cm, width = 8cm]{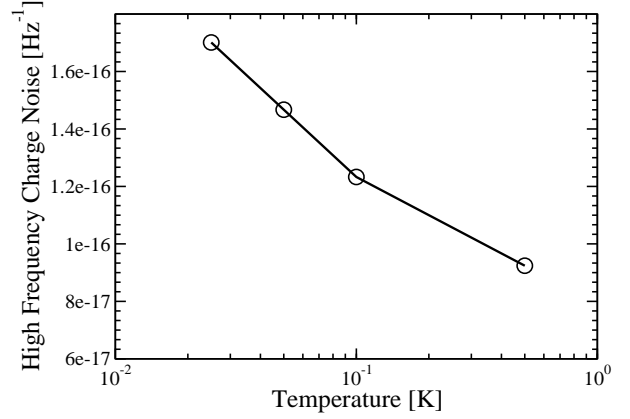}
\caption{\label{fig8} Plot of the high frequency charge noise 
$S_Q(\omega)/e^2$ averaged over the TLS ensemble
versus temperature. Notice that the noise decreases
gradually with increasing temperature. This plot was obtained for
$\omega=10$ GHz, a driving frequency of $\Omega=10$ GHz,
and $I/I_c = 10^7$.}
\end{figure}

Regarding the temperature
dependence, we note that while the low frequency $1/f$ noise is
proportional to temperature, the high frequency white noise 
decreases gradually with increasing temperature as shown in Fig.~\ref{fig8}.
To understand this temperature dependence, note that
at high frequencies the resonant terms dominate. These are the terms 
in Eq.~(\ref{manyTLS}) with $(\omega\pm E/\hbar)$ in the denominator. 
The dominant contribution to the integral occurs at
resonance ($\omega=E/\hbar$) where the temperature dependence of the integrand
goes as $T_2\sim 1/T$ and decreases as the temperature increases. 
However, this decrease is much stronger than seen in the ensemble averaged
noise shown in Fig.~\ref{fig8}. This may be because in obtaining the
ensemble averaged noise, one adds terms away from resonance which have the
opposite trend and increase with increasing temperature as $T_2^{-1}\sim T$.

\begin{figure}
\vspace{0.5cm}
\includegraphics[height = 5.5cm, width = 8cm]{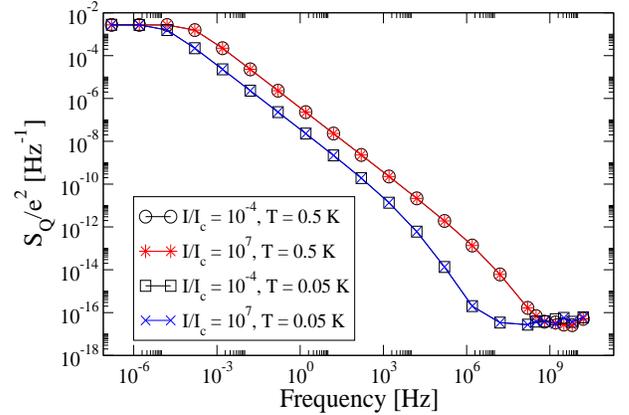}
\caption{\label{fig9} (Color online) 
Log-log plot of the charge noise $S_Q(\omega)/e^2$ 
averaged over the TLS ensemble versus frequency. 
Notice that the results for large and small 
values of $I/I_c$ overlap. We have considered two 
temperatures, 0.5 K and 0.05, respectively, and the driving frequency 
was $\Omega=10$ GHz.}
\end{figure}

In Fig.~\ref{fig9} we show our cumulative numerical results for 
the charge noise spectrum at both low and high frequencies. 
As we mentioned previously, there is no noticeable dependence on the 
electric field intensity ratio $I/I_c$ at either low or high frequencies. 

Our result of white noise at high frequencies disagrees with the
experiments by Astafiev {\it et al.} \cite{Astafiev2004} who concluded that
the noise increases linearly with frequency. However, we caution that
the experiments were done under different conditions from the calculation.
The experimentalists \cite{Astafiev2004} applied dc pulses with Fourier components 
up to a few GHz, possibly saturating TLS with energy splittings in this 
frequency range. They relied on a Landau-Zener transition to excite the qubit 
which had a much larger energy splitting, ranging up to 100 GHz, and measured
the decay rate $\Gamma_1$ of the qubit. Then they used $\Gamma_1$ to
deduce the charge noise spectrum at frequencies equal to the qubit splitting 
by using a formula \cite{Schoelkopf02} derived assuming a stationary state. 
It is not clear whether it is valid to assume
a stationary state in the presence of dc pulses which lasted for a 
time ($\sim$ 100 ps) comparable to the lifetime of the qubit.

In contrast, in calculating noise spectra, we make the customary
assumption of stationarity. We relate the 
charge noise spectra to the response to an ac drive for a broad range of
frequencies. AC driving of qubits have been used in both
theoretical \cite{Shnirman97,You03,Ashhab07} and experimental studies 
\cite{Nakamura01,Oliver05,Sillanpaa06,Berns06} of qubits. 
It would be interesting to measure the frequency and
temperature dependence of the charge noise in Josephson qubits
in the presence of ac driving since no such measurements have been done.

\section{Summary}
\label{Concl}
To summarize, we have studied the effect of the saturation
of TLS by electromagnetic waves on qubit charge noise. 
Using the standard theory of two level systems with a flat 
density of states, we find that the charge noise at low frequencies 
is $1/f$ noise and is insensitive to the saturation of the two level 
systems. In addition the low frequency charge noise increases
linearly with temperature. As one approaches high frequencies, the charge 
noise plateaus to white noise with a very weak dependence 
on the driving frequency $\Omega$ and the ratio $I/I_c$. 
We found that the high frequency 
charge noise decreases slightly with increasing temperature. 

Finally we note that while we have been considering a Josephson junction qubit,
our results on charge and polarization noise have not relied on the 
superconducting properties of the qubit. So our results are much more general
and pertain to the charge noise produced by fluctuating TLS in a 
capacitor or substrate subject to a driving ac electric field in steady state.

This work was supported by the Disruptive Technology Office
under grant W911NF-04-1-0204, and by DOE grant DE-FG02-04ER46107.

$\dagger$Current address: Radiation Physics Division,
Department of Radiation Oncology,
Stanford University, Stanford, CA 94305-5847

\section{Appendix: Derivation of the Charge Induced on the Electrodes by a Dipole}
In this appendix we derive Eq.~(\ref{eq:Q}) which gives the magnitude of 
the charge induced on the electrodes by a dipole Eq.~(\ref{eq:Q}).
Rather than use image charges which leads to an infinite sum, 
we follow Purcell.\cite{Purcell1965} We start by considering the
simple problem of a charge $q$ between two conducting plates connected by a wire
so that they are at the same electric potential. The plates are parallel
to the $xy$ plane and separated by a distance $L$. The sum of the induced
charges is $-q$. Let $d_{u}$ ($d_{\ell}$) be the perpendicular distance
from the charge $q$ to the upper (lower) plate. Let $-q_{u}$ ($-q_{\ell}$)
be the charge induced on the upper (lower) plate so that $-q_{u}-q_{\ell}=-q$.
Notice that if the charge between the plates is doubled to $2q$ such
that $d_{u}$ and $d_{\ell}$ stay the same as before, 
the ratio $q_{u}/q_{\ell}$ stays
the same even though the total induced charge is $-2q$. So let us replace
$q$ by a conducting plate with charge density $\sigma$ while still
maintaining the distances $d_{u}$ and $d_{\ell}$. The total induced 
charge density is $-2\sigma=-\sigma_{u}-\sigma_{\ell}$ where
$-\sigma_{u}$ and $-\sigma_{\ell}$ are the induced charge densities on the
upper and lower plates, respectively. Notice that 
\begin{equation}
\frac{\sigma_{u}}{\sigma_{\ell}}=\frac{q_u}{q_{\ell}}
\label{eq:qRatio}
\end{equation}
Using Gauss' Law to find the electric field and the fact that the voltage
difference between the middle plate and the upper plate equals the
voltage difference between the middle plate and the lower plate yields
\begin{equation}
\frac{\sigma_{u}}{\sigma_{\ell}}=\frac{d_{\ell}}{d_u}
\label{eq:sigmaRatio}
\end{equation}

Now we return to the problem of the charge induced on the plates due
a point charge $q$ between the plates.
From Eqs.~(\ref{eq:sigmaRatio}) and (\ref{eq:qRatio}), we obtain
\begin{equation}
q_u=q\frac{d_{\ell}}{L}
\end{equation}
and
\begin{equation}
q_{\ell}=q\frac{d_u}{L}
\end{equation}

Now suppose we have a dipole between the two conducting plates at the 
same potential.  The dipole consists of two equal and 
opposite charges $q_{+}=q$ and $q_{-}=-q$ separated 
by a distance $d$.  The magnitude of the dipole moment is $p=qd$, 
and $\theta$ is the angle between the z-axis and the dipole moment ${\bf p}$.
Let $d_{u+}$ ($d_{\ell +}$) be the distance between $q_{+}=q$ and 
the upper (lower) plate, and let $d_{u-}$ ($d_{\ell -}$) be the distance
between $q_{-}=-q$ and the upper (lower) plate.
Then the charge induced on the lower plate is $-q_{\ell}$ where
\begin{eqnarray}
q_{\ell}&=&\frac{q}{d}\left(d_{u+}-d_{u-}\right)\nonumber\\
&=&-\frac{p\cos\theta}{L}
\end{eqnarray}
Similarly the charge induced on the upper plate is $-q_{u}$ where
\begin{equation}
q_{u}=\frac{p\cos\theta}{L}
\end{equation}
Thus we recover Eq.~(\ref{eq:Q}) for $Q$, the magnitude of the charge
induced on each electrode.

\end{document}